\documentclass[
    aps,
    prb,
    reprint,
    amsmath,
    amssymb,
    superscriptaddress
]{revtex4-2}

\usepackage{graphicx}% Include figure files
\usepackage{dcolumn}% Align table columns on decimal point
\usepackage{bm}% bold math
\usepackage{color}
\usepackage{hyperref}% add hypertext capabilities
\begin{document}
\raggedbottom
\preprint{APS/123-QED}

\title{\textbf{Geometry-controlled Chern Transfer and Flat Band Reconstruction in Distorted Kagome Lattices} 
}% 

\author{Yu Zhu}

%\affiliation{
% First affiliation for this author
%}%
%\affiliation{
 %second institution for this author
%}%

\author{Claudia Felser}
 % \email{Contact author: Claudia.Felser@cpfs.mpg.de}
\affiliation{Max Planck Institute for Chemical Physics of Solids
}%

\author{Xiaolong Feng}%
 \email{Contact author: Xiaolong.Feng@cpfs.mpg.de}
\affiliation{Max Planck Institute for Chemical Physics of Solids
}%


\date{\today}% It is always \today, today,

\begin{abstract}
Flat band formation and nontrivial topology are central manifestations of
kagome electronic structure, yet they are commonly discussed in idealized
lattice geometries. In distorted kagome materials, structural deformation reorganizes electronic propagation pathways, but a microscopic understanding of how such distortions govern flat band dispersion and band topology is still lacking.
Using a distorted kagome tight-binding model with rotated angle dependent long-range hopping and intrinsic spin--orbit coupling, we show that distortion
reconstructs both the dispersion and topology of the kagome band manifold.
The flat band descendant develops distinct bandwidth regimes associated with
a redistribution of its extrema in momentum space. Simultaneously,
symmetry-related band inversions generate quantized Chern number transfer whose
parity is fixed by the multiplicity of the touching points, thereby
determining the gap-resolved \(\mathbb{Z}_2\) topology. The accompanying
Berry curvature evolution produces characteristic anomalous Hall and
Nernst responses. These results identify geometric deformation as a common
microscopic origin of flat band and topological reconstruction in kagome
systems.
\end{abstract}

                              %display desred
\maketitle

%\tableofcontents

\section{Introduction}

Flat electronic bands provide a favorable setting for correlated and
topological quantum phenomena, as the reduced kinetic-energy scale enhances
the effects of electronic interactions and quantum geometry
\cite{Checkelsky2024FlatBands,Sun2011FlatTopology,Tang2011FQH}.
The kagome lattice constitutes a canonical platform for such physics: destructive interference among electronic hopping paths generates compact
localized states and the characteristic kagome flat band
\cite{Mielke1991LineGraph,Bergman2008BandTouching,Leykam2018FlatBands}, while the accompanying dispersive bands host Dirac crossings and van Hove
singularities
\cite{DiSante2026Kagome,Yin2022Kagome,Kang2020FeSn}. Spin--orbit coupling can then open gaps at relevant band crossings and
induce nontrivial band topology
\cite{KaneMele2005QSH,GuoFranz2009KagomeTI,
Ohgushi2000KagomeHall,Sun2011FlatTopology}.
These characteristic electronic features have been observed across a broad
range of kagome materials, including flat or weakly dispersive bands in
$\mathrm{Fe_3Sn_2}$, FeSn, and CoSn
\cite{Lin2018Fe3Sn2,Ye2018Fe3Sn2,Kang2020FeSn,Kang2020CoSn},
as well as spin--orbit-driven band reconstruction and finite Berry curvature in spin--orbit-coupled kagome systems
\cite{Yin2018SpinOrbit,Yin2019NegativeFlatBand,DiSante2023SpinBerry}.

Moreover, the interference mechanism underlying kagome electronic structure is
intrinsically sensitive to lattice geometry. Structural distortion changes
orbital overlap and reorganizes the hopping network, thereby modifying the
dispersion and band hybridization. This consideration is
particularly relevant to materials in which distorted kagome geometries occur
intrinsically
\cite{Zhao2020HoAgGe,Huang2023CrRhAs,Sinha2021Twisting,
Ortiz2023LnTi3Bi4,Hu2024DistortedTiKagome,Mondal2025NdTi3Bi4}.
Representative examples include HoAgGe and CrRhAs, both of which crystallize
in the hexagonal space group $P\bar{6}2m$ and realize structurally distorted
kagome networks
\cite{Zhao2020HoAgGe,Huang2023CrRhAs}.
The electronic consequences of lattice deformation are evident in the
Berry-curvature-related anomalous Hall response observed in distorted HoAgGe
\cite{Roychowdhury2024HoAgGeAHE} and in theoretical studies demonstrating
pronounced strain-induced modifications of kagome band dispersion and
topological phase boundaries
\cite{Lima2023Strain,Lima2026Topology}.

Despite these developments, the microscopic origin of flat band and
topological reconstruction in distorted kagome systems remains insufficiently
understood. In realistic compounds, structural effects are intertwined with
multiorbital hybridization, crystal-field splitting, magnetic order, and
interlayer hopping, obscuring the intrinsic role of kagome distortion
\cite{Liu2020OrbitalSelectiveCoSn,Okamoto2022MultiorbitalKagome,
Bose2025CoupledKagome,DiSante2026Kagome}. A minimal theoretical description is therefore essential for disentangling
geometric effects for electronic bands and topology.

Here, we propose a tight binding description of distorted kagome lattice with rotated angle dependent long-range hopping and intrinsic spin--orbit coupling. The model isolates how structural deformation reorganizes the electronic pathways governing residual flat band dispersion and topological band reconstruction. The analysis establishes a systematic reconstruction of the kagome flat band manifold, reveals gap-resolved \(\mathbb{Z}_2\) topological transitions governed by symmetry-controlled Chern number transfer, and connects the accompanying redistribution of Berry curvature to anomalous transverse transport. The framework therefore provides a bridge between the fundamental interference
physics of kagome bands and the structurally distorted kagome networks
realized in quantum materials.
\section{Model and Distorted Kagome Lattice}
\label{sec:model}

\subsection{Distorted kagome lattice}
\label{subsec:geometry}

\begin{figure*}
    \centering
    \includegraphics[width=\linewidth]{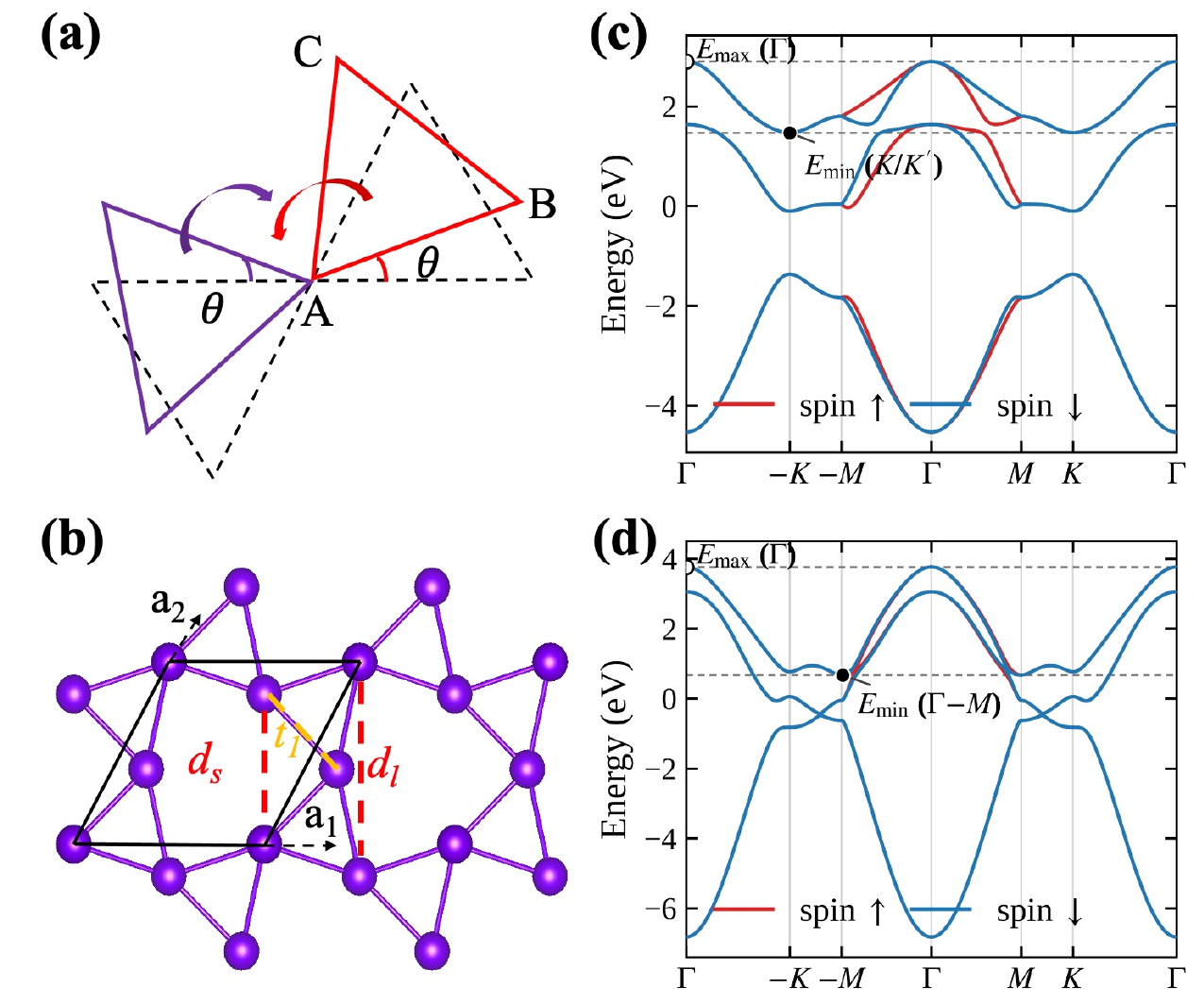}
    \caption{Distorted kagome lattice and representative electronic band structures.
(a) Triangle-rotation geometry characterized by the structural angle
$\theta$, with the regular kagome configuration indicated by dashed lines.
(b) Distorted kagome network showing the nearest-neighbor hopping $t_1$
and the two inequivalent NNN distances, $d_s$ and $d_l$, generated by the
structural distortion.
(c,d) Representative band structures illustrating distinct momentum-space configurations of the flat band descendant. In (c), at $t_2=0.1$ and
$\theta=18^\circ$,
the band maximum remains at $\Gamma$ while the minimum occurs at the
$K/K'$ valleys; in (d), at $t_2=0.7$ and $\theta=3^\circ$, the maximum remains at $\Gamma$ whereas the
minimum is located along the $\Gamma$--$M$ direction. Red and blue curves
denote the spin-up and spin-down sectors, respectively.
    }
    \label{fig:structure}
\end{figure*}

A two-dimensional kagome lattice with three sublattices, $A$, $B$, and $C$,
per unit cell is considered. Motivated by structurally distorted kagome
materials
\cite{Zhao2020HoAgGe,Huang2023CrRhAs,Sinha2021Twisting},
the lattice is deformed by a periodic rotation of neighboring triangular
units, characterized by the structural angle $\theta$, as illustrated in
Fig.~\ref{fig:structure}(a). The regular kagome lattice is recovered at
$\theta=0$.

The deformation preserves the triangular Bravais lattice while changing the
relative positions of the three kagome sublattices within the unit cell.
The neareset-neighbour (NN) bond length $d_0$ is unchanged, so the reference NN hopping $t_1$
remains fixed. By contrast, the next-nearest-neighbour (NNN) bonds split into two
symmetry-related classes (in Fig.~\ref{fig:structure}(b)): a shorter distance $d_{\mathrm{s}}$ and a longer
distance $d_{\mathrm{l}}$, given by
$d_{\mathrm{s}}(\theta)=2d_0
\cos\left(\frac{\pi}{6}+\theta\right)$,
$
d_{\mathrm{l}}(\theta)=2d_0
\cos\left(\frac{\pi}{6}-\theta\right).
\label{eq:NNN_distances}
$
At $\theta=0$, the two become equivalent and recover the NNN distance of the
regular kagome lattice,
$d_{\mathrm{NNN},0}=\sqrt{3}d_0$.

The distortion consequently reorganizes the longer-range hopping network.
Following conventional bond-length-dependent tight-binding
parametrizations
\cite{Pereira2009StrainTB,Lima2023Strain}, the NNN hopping on a bond $ij$
is taken as
$
t_{2,ij}(\theta)
=
t_2
\exp\left[
-\beta_2
\left(
\frac{d_{ij}(\theta)}{d_{\mathrm{NNN},0}}-1
\right)
\right].
\label{eq:t2_distance}
$
Thus, $t_2$ fixes the overall NNN hopping scale, whereas $\beta_2$ determines
how strongly that hopping responds to the distortion-induced change in bond
length. For consistency within the minimal model, an analogous
distance dependence is adopted phenomenologically for the intrinsic intrinsic Kane--Mele-type spin--orbit coupling (SOC),
$
\lambda_{ij}(\theta)
=
\lambda_0
\exp\left[
-\beta_{\mathrm{SOC}}
\left(
\frac{d_{ij}(\theta)}{d_{\mathrm{NNN},0}}-1
\right)
\right],
\label{eq:soc_distance}
$
where $\lambda_0$ is the SOC strength of the undistorted reference lattice.
At $\theta=0$, these expressions reduce to
$t_{2,ij}=t_2$ and $\lambda_{ij}=\lambda_0$ for all symmetry-equivalent NNN
bonds.

The role of the model parameters is therefore separated naturally:
$t_1$ sets the energy scale, $t_2$ controls the overall strength of NNN
electronic propagation, and $\lambda_0$ sets the intrinsic SOC scale,
whereas $\theta$ redistributes these longer-range couplings through the
bond-dependent factors governed by $\beta_2$ and $\beta_{\mathrm{SOC}}$.
This distinction is central to separating the effects of the overall NNN
coupling strength from those arising specifically from lattice distortion.

The triangle rotation breaks $C_2$, $C_6$ symmetry of the regular kagome
lattice while preserving the threefold rotation $C_3$ and three vertical
mirror operations. Within an individual conserved-$s_z$ sector, the
relevant mirror-related constraint is represented by the antiunitary
operation $m\mathcal{T}$. Together with $C_3$ symmetry, these residual symmetries determine the multiplicity of the
band touchings discussed in Sec.~\ref{subsec:band touchings}: $C_3$ generates
three symmetry-related crossings along the $\Gamma$--$M$ directions, whereas
$m\mathcal{T}$ relates the $K$ and $K'$ valleys.

\subsection{Tight binding model}
\label{subsec:Hamiltonian}

The electronic degrees of freedom are described within a single-orbital tight binding framework containing NN and NNN hopping, SOC\cite{KaneMele2005QSH,GuoFranz2009KagomeTI}, and an optional
out-of-plane Zeeman term,
\begin{equation}
\begin{aligned}
H ={}&
t_1\sum_{\langle ij\rangle,s} c_{is}^{\dagger}c_{js}
+\sum_{\langle\!\langle ij\rangle\!\rangle,s}
t_{2,ij}c_{is}^{\dagger}c_{js}
\nonumber\\
&+i\sum_{\langle\!\langle ij\rangle\!\rangle}
\lambda_{ij}\nu_{ij}c_i^{\dagger}s_zc_j
+\Delta_Z\sum_i c_i^{\dagger}s_zc_i .
\end{aligned}
\end{equation}
Here, $t_1$ denotes the NN hopping amplitude, $t_{2,ij}$ the
bond-dependent NNN hopping, and $\lambda_{ij}$ the intrinsic SOC associated
with an NNN hopping path. The factor $\nu_{ij}=\pm1$ specifies the
orientation of the Kane--Mele SOC path, while $\Delta_Z$ denotes the
out-of-plane Zeeman energy.

As we discussed in the section\ref{subsec:geometry}, the NN hopping is chosen as the reference energy scale, $t_1=1 eV$.
The parameter $t_2$ denotes the NNN hopping amplitude of the undistorted
reference lattice and is varied to control the overall strength of
longer-range electronic propagation. The intrinsic SOC scale is fixed at
$\lambda_0=0.1$. Structural modulation of the NNN hopping and SOC is
described by the dimensionless parameters $\beta_2=3$ and
$\beta_{\mathrm{SOC}}=5$, respectively. These parameters quantify the
sensitivity of the corresponding couplings to changes in NNN bond length.

In our model with kane-mele spin orbital coupling effect, the out-of-plane spin component $s_z$ is
conserved
\cite{KaneMele2005QSH,KaneMele2005Z2,GuoFranz2009KagomeTI}. The Bloch Hamiltonian therefore separates into two independent
three-band spin blocks,
$
H(\mathbf{k})
=
H_{\uparrow}(\mathbf{k})
\oplus
H_{\downarrow}(\mathbf{k}).
\label{eq:blockHamiltonian}
$
For $\Delta_Z=0$, the system preserves time-reversal symmetry,
$
H_{\downarrow}(\mathbf{k})
=
H_{\uparrow}^{*}(-\mathbf{k})
$, $E_{n\uparrow}(\mathbf{k})
=
E_{n\downarrow}(-\mathbf{k}).
\label{eq:TR_relation}
$
Each spin block contains three bands, which we label in ascending energy
order as Band 1 ($B1$), Band 2 ($B2$), and Band 3 ($B3$). Unless otherwise
specified, the band-resolved Chern numbers and band-touching processes
discussed below refer to a single spin block

The topological analysis is performed at $\Delta_Z=0$. A finite Zeeman term
is introduced only for the transverse-transport calculations, where it
shifts the two conserved-spin sectors oppositely in energy without modifying
their eigenvectors, Berry curvature, or individual-band Chern numbers.
\section{Results}
\label{sec:results}
\subsection{Distortion-driven reconstruction of the kagome flat band}
\begin{figure}
    \centering
    \includegraphics[width=\linewidth,
    trim={0cm 0.2cm 0cm 0cm},
        clip]{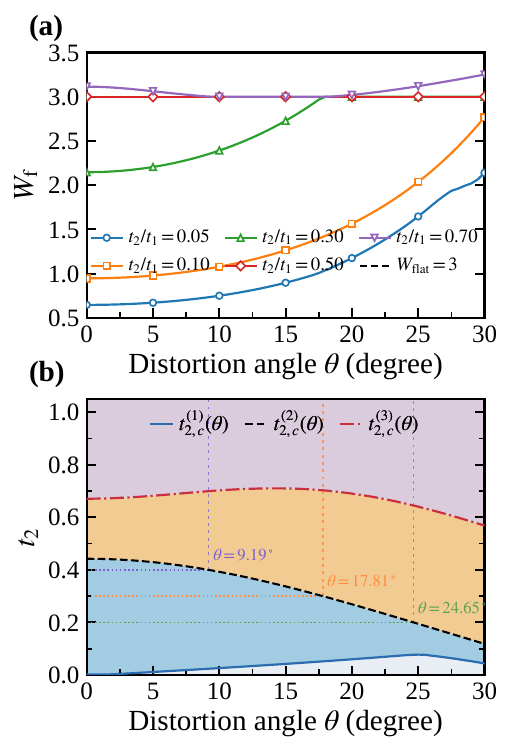}
    \caption{Distortion-driven reconstruction of the kagome flat band.(a) flat band width $W_{\mathrm{flat}}$ as a function of distortion angle
$\theta$ for representative values of $t_2$. The symbols denote the
full-Brillouin-zone numerical results, while the solid curves follow the
corresponding analytical bandwidth expressions. The horizontal dashed line
marks $W_{\mathrm{flat}}=3$.
(b) Bandwidth phase diagram in the $(\theta,t_2)$ plane. The blue and
red boundaries, $t_{2,c}^{(1)}(\theta)$ and
$t_{2,c}^{(3)}(\theta)$, delimit the parameter range in which the
flat band minimum is located at the $K/K'$ valleys, while the black dashed
curve $t_{2,c}^{(2)}(\theta)$ separates the two analytically distinct
$K/K'$-minimum regimes. The four regions correspond to different
momentum-space locations and energetic orderings of the flat band extrema.
The evolution of the global minimum follows
$\Gamma$--$M\rightarrow K/K'\rightarrow\Gamma$--$M$, whereas the band
maximum remains at $\Gamma$.
        }
    \label{fig:flatband_evolution}
\end{figure}
Distortion by triangle rotation reconstructs the flat band through
simultaneous changes in its bandwidth and momentum extrema. This behavior is directly illustrated by the representative
band structures. For $t_2=0.1$ and
$\theta=18^\circ$ in FIG.~\ref{fig:structure}(c), the flat band maximum is located at
$\Gamma$, whereas the minimum occurs at the $K/K'$ valleys. In contrast,
for $t_2=0.7$ and $\theta=3^\circ$ in FIG.~\ref{fig:structure}(d), the maximum remains at
$\Gamma$ while the minimum shifts to the $\Gamma$--$M$ direction.
These two cases demonstrate that the evolution of the nominally flat band
involves not only a change in dispersion, but also a relocation of its
band-edge momentum. 

Here, to quantify this evolution, the flat band width can be defined as
$W_{\mathrm{f}}=\max_{\mathbf{k}}E_3(\mathbf{k})
-\min_{\mathbf{k}}E_3(\mathbf{k})$, where $E_3$ denotes the energy of flat band within single spin channel. As shown in FIG.~\ref{fig:flatband_evolution}(a), the bandwidth is controlled jointly
by the distortion angle $\theta$ and the NNN hopping amplitude $t_2$. Their interplay produces a nonmonotonic reconstruction of the flat band dispersion rather than a simple broadening with increasing distortion.
A full-Brillouin-zone analysis reveals that this reconstruction is governed
by a migration of flat band minimum in momentum space, while the maximum
remains pinned at $\Gamma$ throughout the parameter range considered. Over an intermediate range of $t_2$, it is stabilized at the $K/K'$ valleys. But at
small and large $t_2$, the minimum lies at an off-symmetry momentum along $\Gamma$--$M$ directions. The bandwidth can therefore be expressed as

\begin{equation}
W_{\mathrm{f}}=
\begin{cases}
E_{\Gamma}-E_{\min}^{\Gamma M},
& t_2<t_{2,c}^{(1)}, \\[3pt]

6t_2F_{\beta_2}+2\sqrt{3}\lambda_0F_{\beta_{\mathrm{SOC}}},
& t_{2,c}^{(1)}\le t_2\le t_{2,c}^{(2)}, \\[3pt]

3,
& t_{2,c}^{(2)}<t_2\le t_{2,c}^{(3)}, \\[3pt]

E_{\Gamma}-E_{\min}^{\Gamma M},
& t_2>t_{2,c}^{(3)}.
\end{cases}
\end{equation}

Here, $E_{\min}^{\Gamma M}$ denotes the minimum of the flat band dispersion
along $\Gamma$--$M$ directions.
The three critical NNN hopping scales,
$t_{2,c}^{(1)}$, $t_{2,c}^{(2)}$, and
$t_{2,c}^{(3)}$, separate the four bandwidth regimes.
The first and third boundaries delimit the interval in which the global
flat-band minimum is located at the $K/K'$ valleys, whereas the intermediate
boundary is determined by the crossover between the two $\Gamma/K$ bandwidth
branches,
$
t_{2,c}^{(2)}
=
\frac{
3t_1-2\sqrt{3}\lambda_0
F_{\beta_{\mathrm{SOC}}}
}{
6F_{\beta_{t_2}}
},
$
where
$
F_{\beta}
=
e^{\beta(1-\cos\theta)}
\cosh\!\left[
\frac{\beta}{\sqrt{3}}\sin\theta
\right].
$
The resulting $(\theta,t_2)$ phase diagram, shown in FIG.~\ref{fig:flatband_evolution}(b), separates
four distinct bandwidth regimes. Their boundaries do not merely mark changes
in the magnitude of $W_{\mathrm{flat}}$, but correspond to changes of the band minimum. The sequence of extrema is
$\Gamma$--$M \rightarrow K/K' \rightarrow \Gamma$--$M$, demonstrating that
the distortion reorganizes the momentum-space structure of the flat band
descendant rather than producing a uniform renormalization of its dispersion.
The bandwidth therefore provides a compact measure of how the overall NNN hopping scale and its distortion-induced redistribution jointly reconstruct the interference-generated kagome band. Finite dispersion and structural sensitivity of flat band are also observed in distorted kagome materials such as FeSn and CoSn
\cite{Kang2020FeSn,Kang2020CoSn}, with related titanium-based kagome systems exhibiting multiple weakly dispersive bands and pronounced momentum-space anisotropy
\cite{Hu2024DistortedTiKagome,Mondal2025NdTi3Bi4}.

\begin{figure*}
    \centering
    \includegraphics[
        width=\textwidth,
        trim={0cm 0.4cm 0cm 0cm},
        clip
    ]{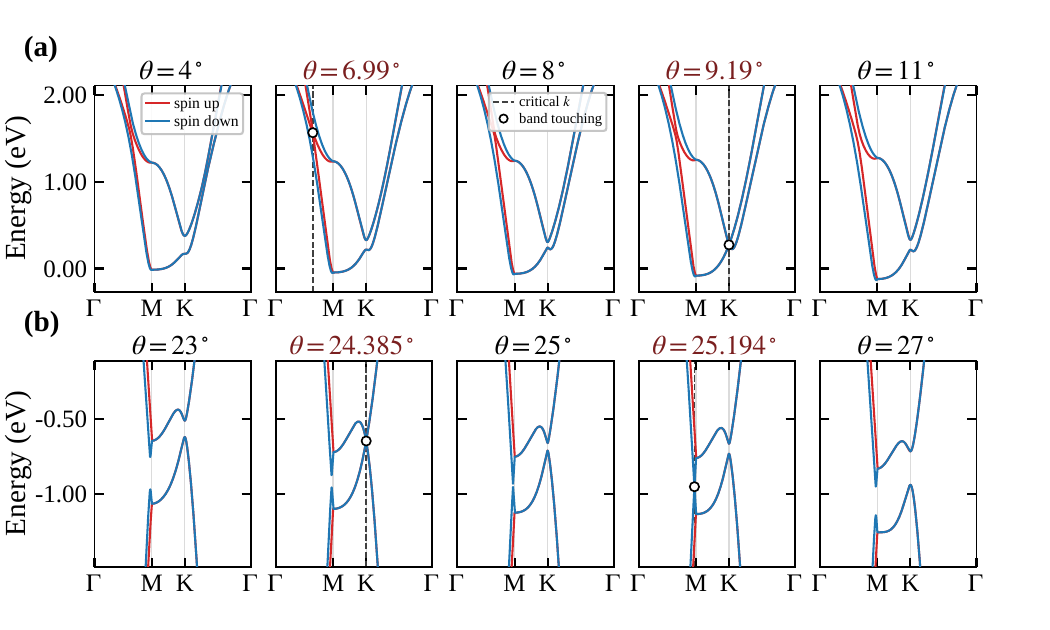}
\caption{
Spin-resolved band evolution across the four distortion-driven topological transitions at $t_2=0.4$.
(a) Evolution of the $B2$--$B3$ bands across the successive gap closings along the symmetry-related $\Gamma$--$M$ directions at $\theta_{1}\simeq6.99^\circ$ and at the $K/K'$ valleys at $\theta_{2}\simeq9.19^\circ$.
(b) Corresponding evolution of the $B1$--$B2$ bands across the $K/K'$ transition at $\theta_{3}\simeq24.385^\circ$ and the subsequent $\Gamma$--$M$ transition at $\theta_{4}\simeq25.194^\circ$.
Red and blue curves denote the spin-up and spin-down sectors, respectively. Vertical dashed lines and open circles indicate the momenta and energies of the critical band touchings. The closing and reopening of the adjacent-band gaps demonstrate the successive distortion-driven band inversions underlying the Chern number transfer.
}
\label{fig:band_touching}
\end{figure*}

Beyond the redistribution of the band extrema, FIG.~\ref{fig:structure}(c)-(d) further reveals that the distortion produces
a pronounced spin-splitting feature along the $\Gamma$--$M$ directions, whereas time-reversal symmetry continues to
relate opposite momenta through
$E_{n\uparrow}(\mathbf{k})=E_{n\downarrow}(-\mathbf{k})$. This behavior follows from the antiunitary operation $m\mathcal{T}$ which acts within an individual spin
sector along the $\Gamma$--$M$ directions and therefore does not enforce
same-$\mathbf{k}$ degeneracy between opposite spins. With increasing distortion, the spin-resolved bands subsequently undergo
a sequence of closing and reopening processes between adjacent bands.
These gap reconstructions indicate that the structural deformation affects
not only the dispersion of the flat band descendant but also the
organization of the neighboring band manifolds, motivating the analysis of
their topological evolution in the following section.

\subsection{Symmetry-controlled Chern transfer and Z2 topology}
\label{subsec:band touchings}

\begin{figure*}
    
    \centering
    \includegraphics[
        width=\textwidth,
        trim={0cm 0.3cm 0cm 0cm},
        clip
    ]{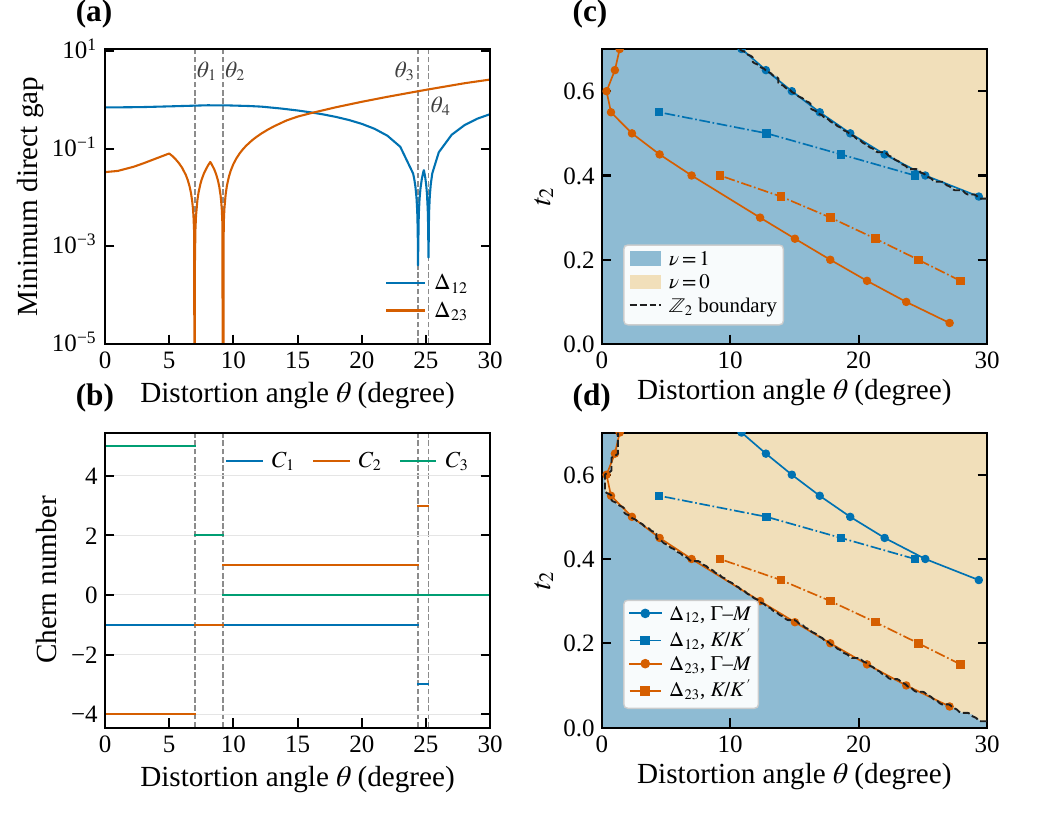}
\caption{
Symmetry-controlled Chern transfer and gap-resolved $\mathbb{Z}_2$ topology.
(a) Minimum direct gaps $\Delta_{12}$ and $\Delta_{23}$ at $t_2=0.4$ as functions of the distortion angle $\theta$. The four gap closings, marked by vertical dashed lines, define the successive band-touching transitions $\theta_{1}$--$\theta_{4}$.
(b) Chern numbers $C_1$, $C_2$, and $C_3$ across the same distortion range, showing the quantized redistribution of Chern number between adjacent bands at each transition.
(c,d) Gap-resolved $\mathbb{Z}_2$ phase diagrams associated with the isolated $B1$ manifold and the $B1$--$B2$ manifold, respectively, in the $(\theta,t_2)$ parameter space. Blue and beige regions denote $\nu=1$ and $\nu=0$, respectively. The overlaid transition lines identify $B1$--$B2$ and $B2$--$B3$ gap closings along the symmetry-related $\Gamma$--$M$ directions and at the $K/K'$ valleys, while the black dashed curves mark the corresponding $\mathbb{Z}_2$ phase boundaries.
}
    \label{fig:z2}
\end{figure*}

The topological reconstruction induced by triangle rotation originates from
a sequence of band inversions between adjacent kagome bands. Along the representative cut $t_2=0.4$, the spin-resolved band structures
in FIG.~\ref{fig:band_touching} directly resolve the sequence of
distortion-driven band inversions. 
The $B2$--$B3$ gap first closes along the three symmetry-related
$\Gamma$--$M$ directions at $\theta_{c1}\simeq6.99^\circ$, followed by a
closing at the $K/K'$ valleys at $\theta_{c2}\simeq9.20^\circ$.
At larger distortion, the $B1$--$B2$ gap closes at $K/K'$ at
$\theta_{c3}\simeq24.39^\circ$ and subsequently along the
$\Gamma$--$M$ directions at $\theta_{c4}\simeq25.19^\circ$.
FIG.~\ref{fig:z2}(a) reveals four successive gap closings with increasing distortion.

The redistribution of topological charge across these inversions is
captured by the band Chern number in FIG.~\ref{fig:z2}(b). With increasing
distortion,
$(C_{B1},C_{B2},C_{B3})$ evolves as
$(1,4,-5)\rightarrow(1,1,-2)\rightarrow(1,-1,0)
\rightarrow(3,-3,0)\rightarrow(0,0,0)$.
At each transition, Chern number is transferred between the two touching
bands while the total Chern number is conserved. The $\Gamma$--$M$
transitions transfer three units of Chern number,
$|\Delta C|=3$, whereas the $K/K'$ transitions transfer two,
$|\Delta C|=2$.
The different transfer magnitudes reflect the symmetry multiplicity of the
underlying band-touching points. A crossing on one $\Gamma$--$M$ direction
is replicated by the threefold rotation $C_3$ onto three symmetry-related
momenta, accounting for the observed threefold transfer. By contrast, the
$K$ and $K'$ crossings form an $m\mathcal{T}$-related valley pair. The corresponding multiplicity is therefore
two, giving an even Chern transfer.

In the time-reversal-symmetric limit considered here, with $s_z$ conserved,
the Hamiltonian decomposes into two time-reversed spin sectors,
$H(\mathbf{k})=
H_{\uparrow}(\mathbf{k})\oplus H_{\downarrow}(\mathbf{k})$,
satisfying
$H_{\downarrow}(\mathbf{k})=
H_{\uparrow}^{*}(-\mathbf{k})$.
The corresponding spin-sector Chern numbers are equal in magnitude and
opposite in sign, so that the total Chern number vanishes. The topology of
the time-reversal-symmetric system is therefore characterized by the
$\mathbb{Z}_2$ invariant, which is given by
the parity of the Chern number of the occupied bands in a single spin sector,
$\nu=C_{\mathrm{occ},\uparrow}\pmod 2$
\cite{KaneMele2005Z2,Sheng2006SpinChern,HasanKane2010TI}.
To identify the isolated band manifolds associated with the two adjacent
gaps, the minimum direct gaps over the full Brillouin zone are defined as
$\Delta_{12}=\min_{\mathbf{k}}
[E_{B2}(\mathbf{k})-E_{B1}(\mathbf{k})]$
and
$\Delta_{23}=\min_{\mathbf{k}}
[E_{B3}(\mathbf{k})-E_{B2}(\mathbf{k})]$.
A finite $\Delta_{12}$ isolates $B1$ from the upper two bands, whereas a
finite $\Delta_{23}$ isolates the $B1$--$B2$ manifold from $B3$.
Accordingly, the corresponding $\mathbb{Z}_2$ invariants are
$\nu_{12}=C_{B1,\uparrow}\pmod 2$
and
$\nu_{23}=(C_{B1,\uparrow}+C_{B2,\uparrow})\pmod 2$.

This formulation makes the connection between Chern transfer and
$\mathbb{Z}_2$ topology explicit. Only a Chern transfer across the
corresponding adjacent-band gap changes the Chern number of the isolated
lower manifold; redistribution entirely within that manifold leaves its
total Chern number unchanged. Moreover, the $\mathbb{Z}_2$ index changes
only when the transferred Chern number is odd. Consequently, the
threefold $\Gamma$--$M$ transition reverses the Chern parity of the
relevant lower manifold, whereas the paired $K/K'$ transition leaves the
parity unchanged. The resulting gap-resolved $\mathbb{Z}_2$ phase diagrams
are shown in FIG.~\ref{fig:z2}(c)-(d),
respectively.

The topological reconstruction is therefore governed by a direct hierarchy:
geometry selects the momentum-space location of the band inversion, while
crystal symmetry fixes the multiplicity of the associated gap closings and,
consequently, the parity of the transferred topological charge. The
$\mathbb{Z}_2$ topology of each isolated band manifold follows from this
symmetry-controlled Chern transfer.

\begin{figure*}
    \centering
    \includegraphics[
        width=\linewidth,
        trim={0cm 0.3cm 0cm 0cm},
        clip
    ]{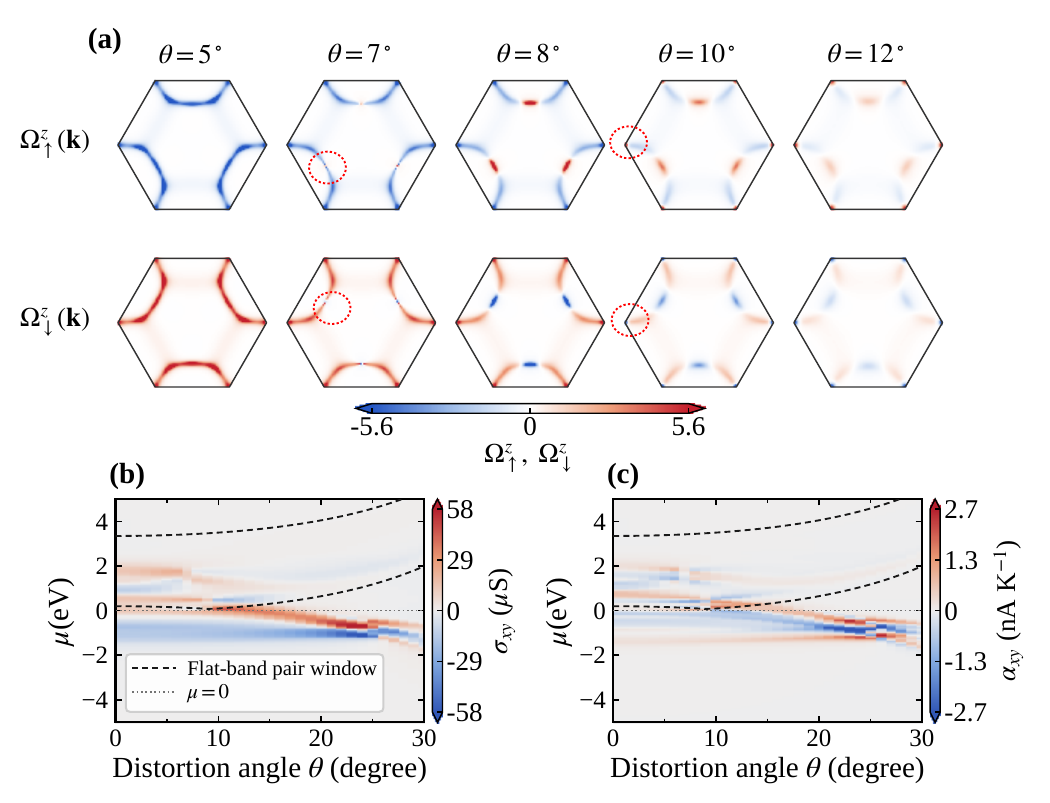}
\caption{
Berry curvature evolution and anomalous transverse transport associated with the distortion-driven band inversions.
(a) Spin-resolved Berry curvature of the flat-band descendant for representative distortion angles at $t_2=0.4$. The two conserved-spin sectors exhibit time-reversed Berry-curvature textures, which are progressively redistributed and concentrated near the momenta involved in the $B2$--$B3$ band inversions. Red dashed circles highlight representative regions associated with the $\Gamma$--$M$ and $K/K'$ transitions.
(b,c) Anomalous Hall conductivity $\sigma_{xy}$ and anomalous Nernst conductivity $\alpha_{xy}$, respectively, over the distortion-angle and chemical-potential parameter space at $T=300$~K. The black dashed curves delimit the energy window of the Zeeman-split upper-band pair, and the horizontal dotted line marks $\mu=0$. A weak out-of-plane Zeeman term is included to lift the compensation between the two time-reversed spin sectors without modifying their eigenstates, Berry curvature, or spin-resolved Chern numbers.
}
\label{fig:transport}
\end{figure*}

\subsection{Berry curvature evolution and transverse response}

The topological reconstruction described above is accompanied by a pronounced
redistribution of Berry curvature, with corresponding consequences for
transverse transport. To examine the anomalous Hall and anomalous Nernst responses of the distorted kagome bands, a weak out-of-plane Zeeman term is
introduced to break time-reversal symmetry. Since $s_z$ remains conserved,
the Zeeman term shifts the two spin sectors oppositely in energy without
modifying their eigenstates. The Berry curvature and spin-resolved Chern
numbers therefore remain unchanged from those of the time-reversal-symmetric
system. The role of the Zeeman term is thus to remove the exact
compensation between the two time-reversed spin sectors and render the
underlying Berry-curvature reconstruction observable in the charge transport
response.

FIG.~\ref{fig:transport}(a) resolves the evolution of the Berry curvature of the flat band descendant across representative distortion angles at
$t_2=0.4$. At small distortion, the Berry-curvature weight is distributed
primarily along extended regions near the Brillouin-zone boundary. Upon approaching the \(B2\)–\(B3\) topological transitions, the Berry curvature develops pronounced localization around the momenta associated with the band inversions, namely along the three symmetry-related \(\Gamma\)–\(M\) directions at \(\theta \approx 6.99^\circ\) and at the \(K/K'\) valleys at \(\theta \approx 9.19^\circ\).
Following the closing and reopening of the gap, both the spatial distribution
and sign structure of the Berry curvature are reorganized. These pronounced
hot spots provide a momentum-space signature of the distortion-driven
$B2$--$B3$ reconstruction identified from the band evolution and Chern-number
transfer in FIG.~\ref{fig:band_touching} and FIG.~\ref{fig:z2}.

The intrinsic anomalous Hall conductivity is given by the occupation-weighted Berry curvature\cite{Nagaosa2010AHE,Xiao2010BerryPhase},
\begin{equation}
\sigma_{xy}(\mu,T)
=
-\frac{e^2}{\hbar}
\sum_{n,s}
\int_{\mathrm{BZ}}
\frac{d^2k}{(2\pi)^2}
\,f_{ns\mathbf{k}}\,
\Omega_{ns}(\mathbf{k}),
\label{eq:ahc}
\end{equation}
,
where $f_{ns\mathbf{k}}$ is the Fermi--Dirac distribution. The corresponding
anomalous Nernst conductivity\cite{Xiao2006ANE} can be written as
\begin{equation}
\alpha_{xy}(\mu,T)
=
\frac{e k_{\mathrm B}}{\hbar}
\sum_{n,s}
\int_{\mathrm{BZ}}
\frac{d^2k}{(2\pi)^2}
\,
\Omega_{ns}(\mathbf{k})
\,s_{ns\mathbf{k}},
\label{eq:anc}
\end{equation}
where
\begin{equation}
s_{ns\mathbf{k}}
=
-f_{ns\mathbf{k}}\ln f_{ns\mathbf{k}}
-
(1-f_{ns\mathbf{k}})
\ln(1-f_{ns\mathbf{k}})
\label{eq:entropy_density}
\end{equation}
is the dimensionless single-state entropy. Whereas $\sigma_{xy}$ reflects the accumulated
Berry curvature of the occupied states, $\alpha_{xy}$ preferentially
weights states within an energy window of order $k_{\mathrm B}T$ around the
chemical potential and is therefore particularly sensitive to rapid
energy-dependent variations of the Berry curvature.

FIG.~\ref{fig:transport}(b)-(c) display the evolution of the anomalous Hall and anomalous Nernst responses, respectively, over the distortion-angle and chemical-potential parameter space. Both quantities exhibit
their strongest variation within the flat band energy window, where the distortion-driven $B2$--$B3$ reconstruction strongly reshapes the Berry-curvature distribution. The response evolves
continuously with distortion but changes most rapidly in the vicinity of the
topological reconstruction. In particular, the sign and magnitude of both $\sigma_{xy}$ and
$\alpha_{xy}$ vary strongly as the chemical potential crosses the reconstructed band manifold, reflecting the transfer of Berry-curvature weight between neighboring bands.
Such Berry-curvature-driven transport is well established in kagome materials. Large intrinsic anomalous Hall conductivities have been reported
in Co$_3$Sn$_2$S$_2$
\cite{Liu2018Co3Sn2S2AHE,Wang2018Co3Sn2S2}, together with a sizable zero-field anomalous Nernst response
\cite{Guin2019Co3Sn2S2ANE}. Related anomalous Hall phenomena occur in
noncollinear kagome magnets such as Mn$_3$Sn
\cite{Nakatsuji2015Mn3Sn} and in the Chern kagome magnet TbMn$_6$Sn$_6$\cite{Yin2020TbMn6Sn6}. Particularly relevant to the present setting,
an enhancement of the anomalous Hall response associated with distorted kagome lattice has been demonstrated in HoAgGe\cite{Roychowdhury2024HoAgGeAHE}. The present results provide complementary
probes of the same underlying geometric reconstruction. The Hall response reflects the occupation-weighted Berry
curvature of the reconstructed bands, whereas the Nernst response is particularly sensitive to Berry-curvature weight in the vicinity of the
chemical potential. Their distortion dependence establishes a direct link between the momentum-space redistribution of Berry curvature and measurable
transverse transport, providing an experimentally accessible signature of the band-topological reconstruction.
\section{Conclusion}
Structural distortion in kagome lattice provides an effective route to engineering band structure and topology through the reorganization of hopping pathways. The flat band develops a strongly momentum-dependent dispersion, with its minimum
shifting between the $\Gamma$--$M$ directions and the
$K/K'$ valleys while the maximum remains at $\Gamma$. The resulting bandwidth
evolution therefore reflects a redistribution of spectral weight in momentum
space rather than a uniform broadening of the flat band manifold.
Moreover, the distortion driven by triangle rotation induces a sequence of band inversions accompanied
by quantized transfer of Chern number between adjacent bands. The topological
outcome is governed by the symmetry multiplicity of the corresponding
gap-closing points: $\Gamma$--$M$ transitions occur at three
symmetry-related momenta and generate an odd Chern transfer, whereas the
$K/K'$ valley pair produces an even transfer. This parity distinction
determines whether the $\mathbb{Z}_2$ invariant associated with a given
isolated band manifold changes. The topological reconstruction is thus set by
the combined action of geometry, which selects the momentum-space location of
the inversion, and crystal symmetry, which fixes the multiplicity of the
associated topological charge transfer.

The band inversions are further accompanied by a pronounced redistribution of
Berry curvature, which is reflected in the anomalous Hall and anomalous Nernst
responses when the exact compensation between time-reversed spin sectors is
removed. These results establish a minimal microscopic description of how
structural distortion reshapes flat band dispersion, band topology, and
Berry-curvature-driven transport in kagome systems, providing a basis for
future material-specific investigations of distorted kagome compounds.

%\begin{acknowledgments}

%\end{acknowledgments}

\bibliography{apssamp}% Produces the bibliography via BibTeX.

@article{Checkelsky2024FlatBands,
  author  = {Checkelsky, Joseph G. and Bernevig, B. Andrei and Coleman, Piers and Si, Qimiao and Paschen, Silke},
  title   = {Flat bands, strange metals and the Kondo effect},
  journal = {Nature Reviews Materials},
  volume  = {9},
  pages   = {509--526},
  year    = {2024},
  doi     = {10.1038/s41578-023-00644-z}
}

@article{Mielke1991LineGraph,
  author  = {Mielke, Andreas},
  title   = {Ferromagnetic ground states for the Hubbard model on line graphs},
  journal = {Journal of Physics A: Mathematical and General},
  volume  = {24},
  number  = {2},
  pages   = {L73--L77},
  year    = {1991},
  doi     = {10.1088/0305-4470/24/2/005}
}

@article{Bergman2008BandTouching,
  author  = {Bergman, Doron L. and Wu, Congjun and Balents, Leon},
  title   = {Band touching from real-space topology in frustrated hopping models},
  journal = {Physical Review B},
  volume  = {78},
  pages   = {125104},
  year    = {2008},
  doi     = {10.1103/PhysRevB.78.125104}
}

@article{Leykam2018FlatBands,
  author  = {Leykam, Daniel and Andreanov, Alexei and Flach, Sergej},
  title   = {Artificial flat band systems: from lattice models to experiments},
  journal = {Advances in Physics: X},
  volume  = {3},
  number  = {1},
  pages   = {1473052},
  year    = {2018},
  doi     = {10.1080/23746149.2018.1473052}
}

@article{DiSante2026Kagome,
  author  = {Di Sante, Domenico and Neupert, Titus and Sangiovanni, Giorgio and Thomale, Ronny and Comin, Riccardo and Checkelsky, Joseph G. and Zeljkovic, Ilija and Wilson, Stephen D.},
  title   = {Kagome metals},
  journal = {Reviews of Modern Physics},
  volume  = {98},
  pages   = {015002},
  year    = {2026},
  doi     = {10.1103/1g9n-wm38}
}

@article{Yin2022Kagome,
  author  = {Yin, Jia-Xin and Lian, Biao and Hasan, M. Zahid},
  title   = {Topological kagome magnets and superconductors},
  journal = {Nature},
  volume  = {612},
  pages   = {647--657},
  year    = {2022},
  doi     = {10.1038/s41586-022-05516-0}
}

@article{KaneMele2005QSH,
  author  = {Kane, C. L. and Mele, E. J.},
  title   = {Quantum Spin Hall Effect in Graphene},
  journal = {Physical Review Letters},
  volume  = {95},
  pages   = {226801},
  year    = {2005},
  doi     = {10.1103/PhysRevLett.95.226801}
}

@article{KaneMele2005Z2,
  author  = {Kane, C. L. and Mele, E. J.},
  title   = {$Z_2$ Topological Order and the Quantum Spin Hall Effect},
  journal = {Physical Review Letters},
  volume  = {95},
  pages   = {146802},
  year    = {2005},
  doi     = {10.1103/PhysRevLett.95.146802}
}

@article{GuoFranz2009KagomeTI,
  author  = {Guo, H.-M. and Franz, M.},
  title   = {Topological insulator on the kagome lattice},
  journal = {Physical Review B},
  volume  = {80},
  pages   = {113102},
  year    = {2009},
  doi     = {10.1103/PhysRevB.80.113102}
}

@article{Ohgushi2000KagomeHall,
  author  = {Ohgushi, Kenya and Murakami, Shuichi and Nagaosa, Naoto},
  title   = {Spin anisotropy and quantum Hall effect in the kagom{\'e} lattice:
             Chiral spin state based on a ferromagnet},
  journal = {Physical Review B},
  volume  = {62},
  pages   = {R6065--R6068},
  year    = {2000},
  doi     = {10.1103/PhysRevB.62.R6065}
}

@article{Tang2011FQH,
  author  = {Tang, Evelyn and Mei, Jia-Wei and Wen, Xiao-Gang},
  title   = {High-Temperature Fractional Quantum Hall States},
  journal = {Physical Review Letters},
  volume  = {106},
  pages   = {236802},
  year    = {2011},
  doi     = {10.1103/PhysRevLett.106.236802}
}

@article{Sun2011FlatTopology,
  author  = {Sun, Kai and Gu, Zhengcheng and Katsura, Hosho and Das Sarma, S.},
  title   = {Nearly Flatbands with Nontrivial Topology},
  journal = {Physical Review Letters},
  volume  = {106},
  pages   = {236803},
  year    = {2011},
  doi     = {10.1103/PhysRevLett.106.236803}
}

@article{Sheng2006SpinChern,
  author  = {Sheng, D. N. and Weng, Z. Y. and Sheng, L. and Haldane, F. D. M.},
  title   = {Quantum Spin-Hall Effect and Topologically Invariant Chern Numbers},
  journal = {Physical Review Letters},
  volume  = {97},
  pages   = {036808},
  year    = {2006},
  doi     = {10.1103/PhysRevLett.97.036808}
}

@article{HasanKane2010TI,
  author  = {Hasan, M. Zahid and Kane, C. L.},
  title   = {Colloquium: Topological insulators},
  journal = {Reviews of Modern Physics},
  volume  = {82},
  pages   = {3045--3067},
  year    = {2010},
  doi     = {10.1103/RevModPhys.82.3045}
}

@article{Lin2018Fe3Sn2,
  author  = {Lin, Zhiyong and Choi, Jin-Ho and Zhang, Qiang and Qin, Wei and Yi, Seho and Wang, Pengdong and Li, Lin and Wang, Yifan and Zhang, Hui and Sun, Zhe and Wei, Laiming and Zhang, Shengbai and Guo, Tengfei and Lu, Qingyou and Cho, Jun-Hyung and Zeng, Changgan and Zhang, Zhenyu},
  title   = {Flatbands and Emergent Ferromagnetic Ordering in {Fe$_3$Sn$_2$} Kagome Lattices},
  journal = {Physical Review Letters},
  volume  = {121},
  pages   = {096401},
  year    = {2018},
  doi     = {10.1103/PhysRevLett.121.096401}
}

@article{Ye2018Fe3Sn2,
  author  = {Ye, Linda and Kang, Mingu and Liu, Junwei and von Cube, Felix and Wicker, Christina R. and Suzuki, Takehito and Jozwiak, Chris and Bostwick, Aaron and Rotenberg, Eli and Bell, David C. and Fu, Liang and Comin, Riccardo and Checkelsky, Joseph G.},
  title   = {Massive Dirac fermions in a ferromagnetic kagome metal},
  journal = {Nature},
  volume  = {555},
  pages   = {638--642},
  year    = {2018},
  doi     = {10.1038/nature25987}
}

@article{Kang2020FeSn,
  author  = {Kang, Mingu and Ye, Linda and Fang, Shiang and You, Jhih-Shih and Levitan, Abe and Han, Minyong and Facio, Jorge I. and Jozwiak, Chris and Bostwick, Aaron and Rotenberg, Eli and others},
  title   = {Dirac fermions and flat bands in the ideal kagome metal {FeSn}},
  journal = {Nature Materials},
  volume  = {19},
  pages   = {163--169},
  year    = {2020},
  doi     = {10.1038/s41563-019-0531-0}
}

@article{Kang2020CoSn,
  author  = {Kang, Mingu and Fang, Shiang and Ye, Linda and Po, Hoi Chun and Denlinger, Jonathan and Jozwiak, Chris and Bostwick, Aaron and Rotenberg, Eli and Kaxiras, Efthimios and Checkelsky, Joseph G. and Comin, Riccardo},
  title   = {Topological flat bands in frustrated kagome lattice {CoSn}},
  journal = {Nature Communications},
  volume  = {11},
  pages   = {4004},
  year    = {2020},
  doi     = {10.1038/s41467-020-17465-1}
}

@article{DiSante2023SpinBerry,
  author  = {Di Sante, Domenico and Bigi, Chiara and Eck, Philipp and Enzner, Stefan and Consiglio, Armando and Pokharel, Ganesh and Carrara, Pietro and Orgiani, Pasquale and Polewczyk, Vincent and Fujii, Jun and others},
  title   = {Flat band separation and robust spin Berry curvature in bilayer kagome metals},
  journal = {Nature Physics},
  volume  = {19},
  pages   = {1135--1142},
  year    = {2023},
  doi     = {10.1038/s41567-023-02053-z}
}

@article{Yin2018SpinOrbit,
  author  = {Yin, Jia-Xin and Zhang, Songtian S. and Chang, Guoqing and others},
  title   = {Giant and anisotropic many-body spin--orbit tunability in a strongly correlated kagome magnet},
  journal = {Nature},
  volume  = {562},
  pages   = {91--95},
  year    = {2018},
  doi     = {10.1038/s41586-018-0502-7}
}

@article{Yin2019NegativeFlatBand,
  author  = {Yin, Jia-Xin and Zhang, Songtian S. and Chang, Guoqing and Wang, Qi and Tsirkin, Stepan S. and Guguchia, Zurab and Lian, Biao and Zhou, Huibin and Jiang, Kun and Belopolski, Ilya and others},
  title   = {Negative flat band magnetism in a spin--orbit-coupled correlated kagome magnet},
  journal = {Nature Physics},
  volume  = {15},
  pages   = {443--448},
  year    = {2019},
  doi     = {10.1038/s41567-019-0426-7}
}

@article{Yin2020TbMn6Sn6,
  author  = {Yin, Jia-Xin and Ma, Wenlong and Cochran, Tyler A. and Xu, Xitong and Zhang, Songtian S. and Tien, Hung-Ju and Shumiya, Nana and Cheng, Guangming and Jiang, Kun and Lian, Biao and others},
  title   = {Quantum-limit Chern topological magnetism in {TbMn$_6$Sn$_6$}},
  journal = {Nature},
  volume  = {583},
  pages   = {533--536},
  year    = {2020},
  doi     = {10.1038/s41586-020-2482-7}
}

@article{Liu2018Co3Sn2S2AHE,
  author  = {Liu, Enke and Sun, Yan and Kumar, Nitesh and Muechler, Lukas and Sun, Aili and Jiao, Lin and Yang, Shuo-Ying and Liu, Defa and Liang, Aiji and Xu, Qiunan and others},
  title   = {Giant anomalous Hall effect in a ferromagnetic kagome-lattice semimetal},
  journal = {Nature Physics},
  volume  = {14},
  pages   = {1125--1131},
  year    = {2018},
  doi     = {10.1038/s41567-018-0234-5}
}

@article{Wang2018Co3Sn2S2,
  author  = {Wang, Qi and Xu, Yuanfeng and Lou, Rui and Liu, Zhonghao and Li, Man and Huang, Yaobo and Shen, Dawei and Weng, Hongming and Wang, Shancai and Lei, Hechang},
  title   = {Large intrinsic anomalous Hall effect in half-metallic ferromagnet {Co$_3$Sn$_2$S$_2$} with magnetic Weyl fermions},
  journal = {Nature Communications},
  volume  = {9},
  pages   = {3681},
  year    = {2018},
  doi     = {10.1038/s41467-018-06088-2}
}

@article{Nakatsuji2015Mn3Sn,
  author  = {Nakatsuji, Satoru and Kiyohara, Naoki and Higo, Tomoya},
  title   = {Large anomalous Hall effect in a non-collinear antiferromagnet at room temperature},
  journal = {Nature},
  volume  = {527},
  pages   = {212--215},
  year    = {2015},
  doi     = {10.1038/nature15723}
}

@article{Zhao2020HoAgGe,
  author  = {Zhao, Kan and Deng, Hao and Chen, Hua and Ross, Kate A. and Pet{\v{r}}{\'i}{\v{c}}ek, V{\'a}clav and G{\"u}nther, Gerrit and Russina, Margarita and Hutanu, Vladimir and Gegenwart, Philipp},
  title   = {Realization of the kagome spin ice state in a frustrated intermetallic compound},
  journal = {Science},
  volume  = {367},
  number  = {6483},
  pages   = {1218--1223},
  year    = {2020},
  doi     = {10.1126/science.aaw1666}
}

@article{Huang2023CrRhAs,
  author  = {Huang, Y. N. and Jeschke, Harald O. and Mazin, Igor I.},
  title   = {{CrRhAs}: a member of a large family of metallic kagome antiferromagnets},
  journal = {npj Quantum Materials},
  volume  = {8},
  pages   = {32},
  year    = {2023},
  doi     = {10.1038/s41535-023-00562-x}
}

@article{Roychowdhury2024HoAgGeAHE,
  author  = {Roychowdhury, Subhajit and Samanta, Kartik and Singh, Sukriti and Schnelle, Walter and Zhang, Yang and Noky, Jonathan and Vergniory, Maia G. and Shekhar, Chandra and Felser, Claudia},
  title   = {Enhancement of the anomalous Hall effect by distorting the Kagome lattice in an antiferromagnetic material},
  journal = {Proceedings of the National Academy of Sciences of the United States of America},
  volume  = {121},
  number  = {30},
  pages   = {e2401970121},
  year    = {2024},
  doi     = {10.1073/pnas.2401970121}
}

@article{Lima2023Strain,
  author  = {Lima, W. P. and da Costa, D. R. and Sena, S. H. R. and Pereira, J. Milton, Jr.},
  title   = {Effects of uniaxial and shear strains on the electronic spectrum of Lieb and kagome lattices},
  journal = {Physical Review B},
  volume  = {108},
  pages   = {125433},
  year    = {2023},
  doi     = {10.1103/PhysRevB.108.125433}
}

@article{Lima2026Topology,
  author  = {Lima, W. P. and Lara, T. F. O. and Nascimento, J. P. G. and Pereira, J. Milton, Jr. and da Costa, D. R.},
  title   = {Topological phase transitions in strained Lieb-kagome lattices},
  journal = {Physical Review B},
  volume  = {114},
  pages   = {045402},
  year    = {2026},
  doi     = {10.1103/3xb5-g78k}
}

@article{Sinha2021Twisting,
  author  = {Sinha, Mekhola and Vivanco, Hector K. and Wan, Cheng and Siegler, Maxime A. and Stewart, Veronica J. and Pogue, Elizabeth A. and Pressley, Lucas A. and Berry, Tanya and Wang, Ziqian and Johnson, Isaac and others},
  title   = {Twisting of 2D Kagom{\'e} Sheets in Layered Intermetallics},
  journal = {ACS Central Science},
  volume  = {7},
  number  = {8},
  pages   = {1381--1390},
  year    = {2021},
  doi     = {10.1021/acscentsci.1c00599}
}

@article{Ortiz2023LnTi3Bi4,
  author  = {Ortiz, Brenden R. and Miao, Hu and Parker, David S. and Yang, F. and Samolyuk, G. D. and Clements, E. M. and Rajapitamahuni, A. and Yilmaz, T. and Vescovo, E. and Yan, J. and others},
  title   = {Evolution of Highly Anisotropic Magnetism in the Titanium-Based Kagome Metals {LnTi$_3$Bi$_4$} ({Ln}: La$\cdots$Gd$^{3+}$, Eu$^{2+}$, Yb$^{2+}$)},
  journal = {Chemistry of Materials},
  volume  = {35},
  number  = {22},
  pages   = {9756--9773},
  year    = {2023},
  doi     = {10.1021/acs.chemmater.3c02289}
}

@article{Hu2024DistortedTiKagome,
  author  = {Hu, Yong and Le, Congcong and Chen, Long and Deng, Hanbin and Zhou, Ying and Plumb, Nicholas C. and Radovic, Milan and Thomale, Ronny and Schnyder, Andreas P. and Yin, Jia-Xin and Wang, Gang and Wu, Xianxin and Shi, Ming},
  title   = {Magnetic coupled electronic landscape in bilayer-distorted titanium-based kagome metals},
  journal = {Physical Review B},
  volume  = {110},
  pages   = {L121114},
  year    = {2024},
  doi     = {10.1103/PhysRevB.110.L121114}
}

@article{Mondal2025NdTi3Bi4,
  author  = {Mondal, Mazharul Islam and Sakhya, Anup Pradhan and Sprague, Milo and Ortiz, Brenden R. and Matzelle, Matthew and Kumay, Arun K. and Seal, Avike and Ghosh, Barun and Bansil, Arun and Neupane, Madhab},
  title   = {Observation of multiple flat bands and van Hove singularities in the distorted kagome metal {NdTi$_3$Bi$_4$}},
  journal = {Physical Review B},
  volume  = {112},
  pages   = {L121104},
  year    = {2025},
  doi     = {10.1103/mhj5-ws5v}
}

@article{Pereira2009StrainTB,
  author  = {Pereira, Vitor M. and Castro Neto, A. H. and Peres, N. M. R.},
  title   = {Tight-binding approach to uniaxial strain in graphene},
  journal = {Physical Review B},
  volume  = {80},
  pages   = {045401},
  year    = {2009},
  doi     = {10.1103/PhysRevB.80.045401}
}

@article{Xiao2010BerryPhase,
  author  = {Xiao, Di and Chang, Ming-Che and Niu, Qian},
  title   = {Berry phase effects on electronic properties},
  journal = {Reviews of Modern Physics},
  volume  = {82},
  pages   = {1959--2007},
  year    = {2010},
  doi     = {10.1103/RevModPhys.82.1959}
}

@article{Nagaosa2010AHE,
  author  = {Nagaosa, Naoto and Sinova, Jairo and Onoda, Shigeki and MacDonald, A. H. and Ong, N. P.},
  title   = {Anomalous Hall effect},
  journal = {Reviews of Modern Physics},
  volume  = {82},
  pages   = {1539--1592},
  year    = {2010},
  doi     = {10.1103/RevModPhys.82.1539}
}

@article{Xiao2006ANE,
  author  = {Xiao, Di and Yao, Yugui and Fang, Zhong and Niu, Qian},
  title   = {Berry-Phase Effect in Anomalous Thermoelectric Transport},
  journal = {Physical Review Letters},
  volume  = {97},
  pages   = {026603},
  year    = {2006},
  doi     = {10.1103/PhysRevLett.97.026603}
}

@article{Guin2019Co3Sn2S2ANE,
  author  = {Guin, Satya N. and Vir, Praveen and Zhang, Yang and Kumar, Nitesh and Watzman, Sarah J. and Fu, Chenguang and Liu, Enke and Manna, Kaustuv and Schnelle, Walter and Gooth, Johannes and Shekhar, Chandra and Sun, Yan and Felser, Claudia},
  title   = {Zero-Field Nernst Effect in a Ferromagnetic Kagome-Lattice Weyl-Semimetal {Co$_3$Sn$_2$S$_2$}},
  journal = {Advanced Materials},
  volume  = {31},
  pages   = {1806622},
  year    = {2019},
  doi     = {10.1002/adma.201806622}
}

@article{Liu2020OrbitalSelectiveCoSn,
  author  = {Liu, Zhonghao and Li, Man and Wang, Qi and Wang, Guangwei and Wen, Chenhaoping and Jiang, Kun and Lu, Xiangle and Yan, Shichao and Huang, Yaobo and Shen, Dawei and Yin, Jia-Xin and Wang, Ziqiang and Yin, Zhiping and Lei, Hechang and Wang, Shancai},
  title   = {Orbital-selective Dirac fermions and extremely flat bands in frustrated kagome-lattice metal {CoSn}},
  journal = {Nature Communications},
  volume  = {11},
  pages   = {4002},
  year    = {2020},
  doi     = {10.1038/s41467-020-17462-4}
}

@article{Okamoto2022MultiorbitalKagome,
  author  = {Okamoto, Satoshi and Mohanta, Narayan and Dagotto, Elbio and Sheng, D. N.},
  title   = {Topological flat bands in a kagome lattice multiorbital system},
  journal = {Communications Physics},
  volume  = {5},
  pages   = {198},
  year    = {2022},
  doi     = {10.1038/s42005-022-00969-1}
}

@article{Bose2025CoupledKagome,
  author  = {Bose, Anumita and Bandyopadhyay, Arka and Narayan, Awadhesh},
  title   = {Origin of flat bands and non-trivial topology in coupled kagome lattices},
  journal = {Communications Physics},
  volume  = {8},
  pages   = {519},
  year    = {2025},
  doi     = {10.1038/s42005-025-02432-3}
}

\end{document}